\documentclass{ws-procs9x6}

\begin{document}
  \newcommand {\nc} {\newcommand}
  \nc {\beq} {\begin{eqnarray}}
  \nc {\eeq} {\nonumber \end{eqnarray}}
  \nc {\eeqn}[1] {\label {#1} \end{eqnarray}}
  \nc {\ve} [1] {\mbox{\boldmath $#1$}}

\title{Are present reaction theories for studying rare isotopes good enough?}

\author{F. M. Nunes$^1$, P. Capel$^2$, R.J. Charity$^3$, A. Deltuva$^4$,
W.Dickhoff$^5$, H. Esbensen$^6$, \\
R.C. Johnson$^7$, N.B. Nguyen$^1$, N.J. Upadhyay$^1$ and S.J. Waldecker$^5$}

\address{$^1$National Superconducting Cyclotron Laboratory and Department of Physics and Astronomy, Michigan State University, East Lansing, MI 48824, USA}

\address{$^2$Helmholtz-Insitut Mainz, Johannes Gutenberg-Universit\"at, D-55128 Mainz, Germany}

\address{$^3$Department of Chemistry, Washington University, St. Louis, Missouri 63130, USA}

\address{$^4$Centro de F\'{\i}sica Nuclear da Universidade de Lisboa, P-1649-003 Lisboa, Portugal}

\address{$^5$Department of Physics, Washington University, St. Louis, Missouri 63130, USA}

\address{$^6$Physics Division, Argonne National Laboratory, Argonne, Illinois, 60439, USA}

\address{$^7$Department of Physics, University of Surrey, Guildford GU2 7XH, United Kingdom}

\begin{abstract}
Rare isotopes are most often studied through nuclear reactions. Nuclear
reactions can be used to obtain detailed structure information but also in
connection to astrophysics to determine specific capture rates. In order
to extract the desired information it is crucial to have a reliable
framework that describes the reaction process accurately. A few recent
developments for transfer and breakup reactions will be presented. These include
recent studies on the reliability of existing theories as well as effort to
reduce the ambiguities in the predicted observables.
\end{abstract}


\bodymatter

\newpage
\section{Introduction}\label{intro}

The study of rare isotopes has challenged many of our traditional concepts in nuclear
physics. New phenomena have been unveiled and, alongside, new approaches have been developed.
An example is the loosely bound nature of many of these exotic nuclei which stimulated the
development of new theoretical frameworks that can handle open systems at the same level as closed systems.
In order to assess the progress in our understanding of nuclei away from stability,
many experiments continue to be performed at the edge of technical capabilities.
Most often these experiments involve reactions of the species of interest with a stable
target. Through reactions, one can probe shell structure, correlations, pairing, the role of the
continuum, etc \cite{review}. One can also extract radiative capture rates relevant for astrophysics \cite{solar}. Ultimately, to obtain a meaningful conclusion from the comparison with the data,
a reliable reaction model is needed.

In a fully microscopic framework, reaction and structure emerge from the same framework.
The efforts on ab-initio reactions have increased significantly over the last decade and resulted
in very important progress (e.g.  \cite{nollet,navratil}). Present implementations do not include
the three-body nucleon force, nor are they able to include three-body dynamics. These are
crucial advances that will increase predictability.
However, ab-initio formulations are limited to light nuclear reactions. For all other cases of
interest, the many-body problem needs to be simplified such that the important dynamics in
the reaction process can be properly accounted for.

Most reaction theories thus rely on the
reduction of the many-body problem to a few-body problem, in which the relevant degrees of
freedom have been preserved. In doing this reduction, one introduces ambiguities, since the
effective interactions between the reduced clusters are not uniquely determined. Also,
in solving the few-body problem, a number of different approximations are applied.
The ambiguities introduced in the reduction of the many-body problem need to be kept
under control and uncertainties in the treatment in the dynamics need to be quantified
in a systematic way. Here we  discuss a variety of recent efforts to quantify the
errors or reduce ambiguities in present treatments.

In Section 2, we focus on transfer reactions, particularly (d,p).
Now that the exact full three-body Faddeev solution for the scattering problem has become
feasible, several benchmarks of existing theories have been recently carried out
\cite{nunes11,nunes11b} (see Section 2.1).
The use of more microscopically informed effective interactions in the description of transfer
reactions has also proven to be beneficial. The example discussed here is the use of the
dispersive optical models to determine all inputs to (d,p) reactions on double
magic nuclei \cite{nguyen11}.

In Section 3. we turn to breakup reaction models. First, we discuss the benchmark of two existing
non-perturbative theories for dealing with breakup on nuclear targets, which includes reactions dominated
by the Coulomb interaction \cite{capel11}. Even if the reaction dynamics is accurately described, the breakup
cross section depends strongly on the core-target interaction. In Section 3.2, we discuss a
new method proposed to reduce the uncertainties introduced exactly by the core-target interaction
when extracting structure information from  breakup \cite{ratio}.

\section{Transfer reactions}

Transfer (d,p)/(p,d)  reactions are the most common tool to study single particle structure.
Generally, from the transfer angular distributions one obtains the angular momentum of the state while 
the normalization at forward angles provides the spectroscopic factor \cite{book}.

\subsection{Testing models}

It has long been understood that in (d,p) reactions, deuteron breakup can play an important role \cite{johnson-ria}.
The adiabatic approach was originally developed within the zero range approximation. However,
recently, a systematic study using the finite range adiabatic model \cite{tandy} concluded
that finite range effects should be properly taken into account, not only in the calculation of
the transfer matrix elements but also in treating the deuteron breakup \cite{nguyen10}.
In this model, the three-body wavefunction, describing $d+A$ scattering, is expanded in Weinberg
states and truncated to first order \cite{tandy}.

The continuum discretized coupled channel method (CDCC) \cite{cdcc} provides a more accurate approach of treating breakup,
in that it does not make the aforementioned truncation. In CDCC the elastic and breakup channels are
fully coupled. To compute the transfer channel, one replaces in the transfer T-matrix,
the exact three-body wavefunction by the CDCC wavefunction.  Until recently,
the error in that approximation has not been quantified.

The most accurate way of solving the three-body problem in the presence of rearrangement channels
is by using an overcomplete basis written in all three Jacobi components. This gives way to
the Faddeev equations which in the few-nucleon field has traditionally been solved as integral
equations in momentum space. Critical for the extension of those techniques into nuclear reactions
with heavier systems was the progress in the treatment of the Coulomb \cite{deltuva}.
\begin{figure}[t]
\begin{center}
\psfig{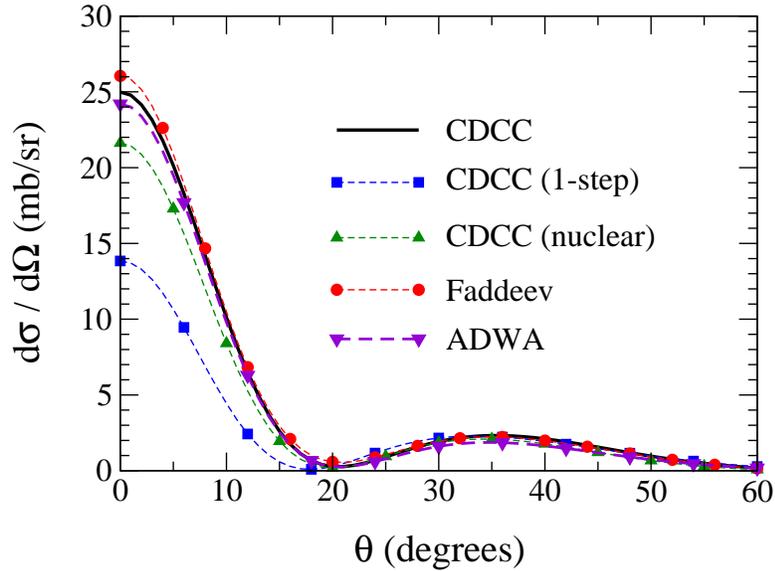}
\end{center}
\caption{Angular distribution for $^{10}$Be(d,p)$^{11}$Be at 21.4 MeV: comparing CDCC (solid) with Faddeev
(filled circles). Also shown are the results ignoring Coulomb (triangles) and neglecting the coupling
to breakup (squares).}
\label{cdcc-fadd}
\end{figure}

Starting from the same Hamiltonian we have performed a systematic comparison of ADWA and Faddeev \cite{nunes11}.
This study includes a range of target masses from $A=10$ to $A=48$, as well as a range of energies
$E \approx 5-40$ A MeV. Results show that around $E=10$ A MeV there is good agreement
between the two approaches. However the agreement deteriorates at higher energies, where ambiguities
in the interactions play a significant role.
A similar study for CDCC versus Faddeev was performed \cite{upadhyay} and qualitatively,
the same conclusions were drawn. 
In Fig. \ref{cdcc-fadd} we show one example studied, the $^{10}$Be(d,p)$^{11}$Be reaction at
21.4 MeV. Here all three methods ADWA, CDCC and Faddeev predict cross sections within a few percent
from each other. Also shown is the importance of Coulomb in determining the theoretical normalization
of angular distribution at forward angles accurately. Because the treatment of Coulomb is an important
limiting factor in extending Faddeev calculations to heavy nuclei, it is useful to understand whether
its contribution is significant. Fig. \ref{cdcc-fadd} illustrates that not including Coulomb,
even for such a light system, introduces an error of $15$\%.
Finally, Fig. \ref{cdcc-fadd} also shows the results obtained when no breakup is considered,
and the transfer process takes place in 1-step. The large difference between CDCC and CDCC(1-step) 
confirm the need to include breakup in this process.

\subsection{Reducing ambiguities}

Important ingredients in predicting the angular distributions following a A(d,p)B reaction
are the optical potentials n-A and p-A. These are usually taken from global parameterizations to elastic scattering
data (e.g. \cite{ch89}). In addition one also needs a description of the final bound state of interest.
For this purpose, we often adjust a real Woods-Saxon potential with standard parameters to reproduce
the separation energy and other basic properties of the nucleus B.

\begin{figure}[t]
\begin{center}
\psfig{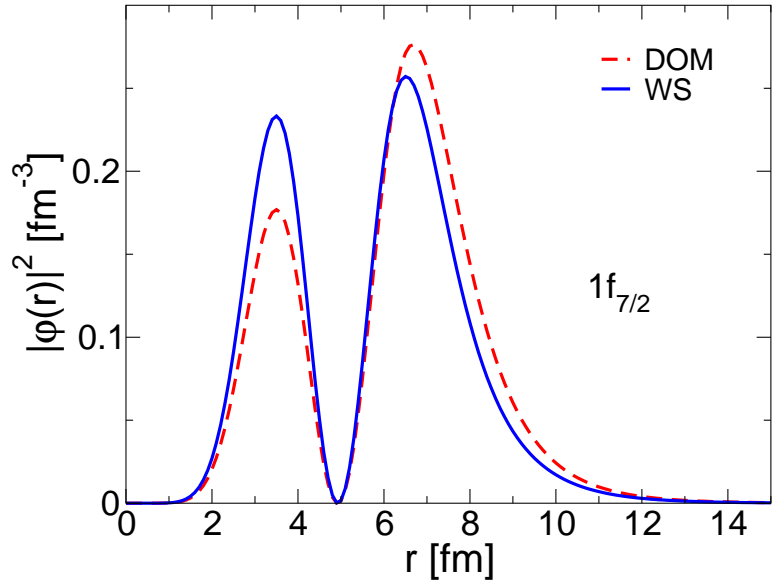}
\end{center}
\caption{Single neutron overlap function for the ground state of $^{133}$Sn obtained from a
Woods Saxon potential and with the local DOM corrected for non-locality.}
\label{dom-overlap}
\end{figure}
\begin{figure}
\begin{center}
\psfig{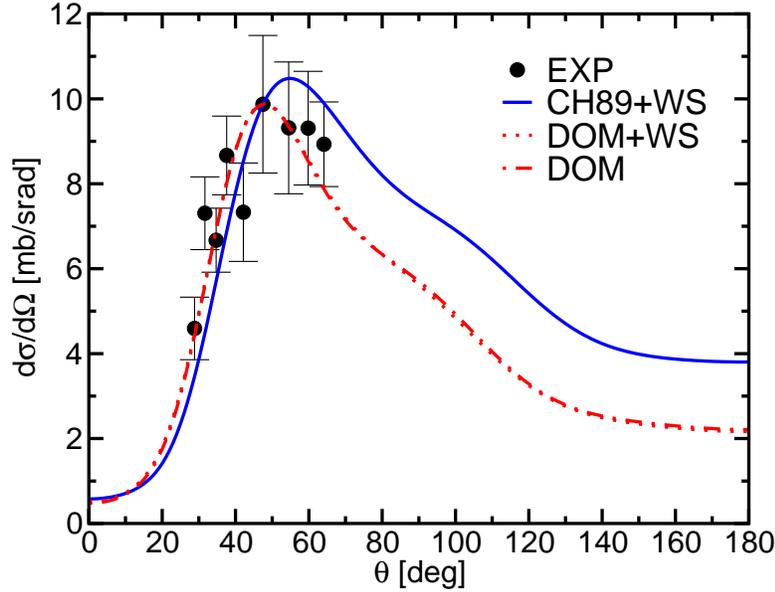}
\end{center}
\caption{Angular distribution for $^{132}$Sn(d,p)$^{133}$Sn at 9.46 MeV}
\label{dom-xs}
\end{figure}
The dispersive optical model (DOM) \cite{mahaux} is based on the microscopic Green's function approach which connects the scattering states and the
bound states through dispersive relations. The immediate advantage is that the data driven extrapolations are guided by microscopic theory. Recent studies using DOM on the Ca isotopes show unusual features in the isospin dependence of the nucleon interaction \cite{mueller11}. This becomes particularly relevant when extrapolating to unknown regions of the nuclear chart.

DOM optical potentials were applied to (d,p) reactions on several doubly magic nuclei \cite{nguyen11}. Therein, transfer calculations were performed within the finite range ADWA. Results show that in most cases, DOM is able to describe the angular distributions as well as the global Chapel Hill parameterization \cite{ch89}, which is commonly used. However the normalization of the cross sections obtained with DOM are reduced and become more in line with results from (e,e'p)  measurements, mostly due to the non-locality included in the
bound state wavefunction.

Another important result has to do with the beam energy dependence in the extracted spectroscopic factors \cite{nguyen11}.
For $^{49}$Ca, while using the standard approach produces a wide range of spectroscopic factors depending on the beam energy,
within DOM, the range of extracted values are much less dependent on the beam energy.

Probably the most interesting result in Nguyen {\it et al.} \cite{nguyen11} concerns the exotic $^{133}$Sn.
In Fig. \ref{dom-overlap} we show the square of the single neutron bound state for $^{133}$Sn in the ground state $| \varphi_n|^2$ as a function of the n-$^{132}$Sn distance (r), calculated within
the standard Woods-Saxon approach and the DOM. Non-locality shifts strength from the inner part of the bound state to the surface. For peripheral reactions, this results in larger cross sections and smaller 
extracted spectroscopic factors.
In Fig. \ref{dom-xs} we show the angular distribution for $^{132}$Sn(d,p)$^{133}$Sn at 9.5 MeV. We plot
the predictions and data from  \cite{jones11} as well as those using DOM input. Results indicate that for this
system DOM provides an improvement in the description of the angular distribution as compared to Chapel-Hill.
This is understood in terms of a more reliable extrapolation to neutron rich systems.

\section{Breakup reactions}

Breakup reactions are a useful tool to study nuclear structure properties but also provide
an indirect method for direct capture of astrophysical relevance.
The relative energy distribution is strongly dependent on the binding energy and the angular momentum, but also to the asymptotic normalization of the overlap function of the initial ground state. It also enables the study of the resonances of the nucleus of interest.
Here we focus on two-body breakup, corresponding to reactions of the type $P+T \rightarrow c+x+T$.

\subsection{Testing models}

As pointed our in Section 2., an exact approach for solving the three-body $c+x+T$ involves solving the Faddeev Equations.
However, because most of the breakup experiments are performed on
nuclear targets with charge $Z>1$, present implementations of the Faddeev equations
in nuclear reactions are not feasible.

The best available method for solving the problem is CDCC \cite{cdcc}. However CDCC is computationally
intensive and generalizations of CDCC to include more complex projectile structure
are often hindered by computational limitations. It is therefore very desirable to have
an alternative method that is equally accurate.
\begin{figure}[t]
\begin{center}
\psfig{file=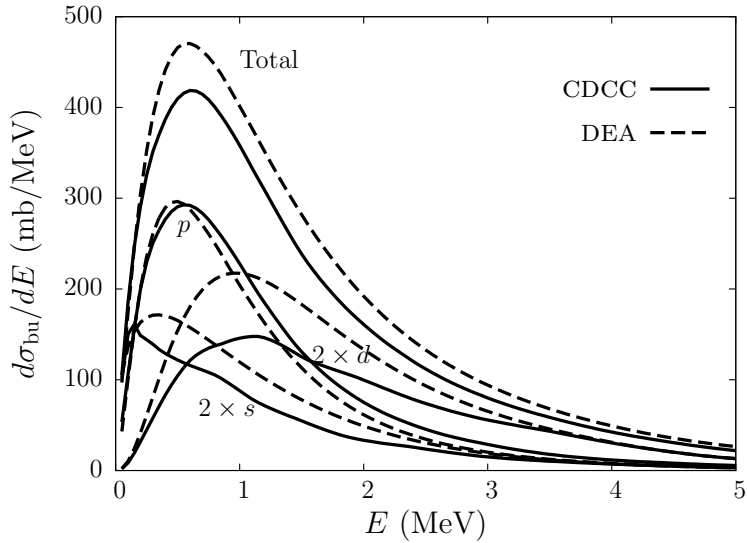,width=4in}
\end{center}
\caption{Relative energy $^{14}$C-n distribution following the breakup of $^{15}$C on Pb at 20 AMeV: comparing the partial wave contributions for CDCC and DEA.}
\label{cdcc-dea20}
\end{figure}
\begin{figure}[h]
\begin{center}
\psfig{file=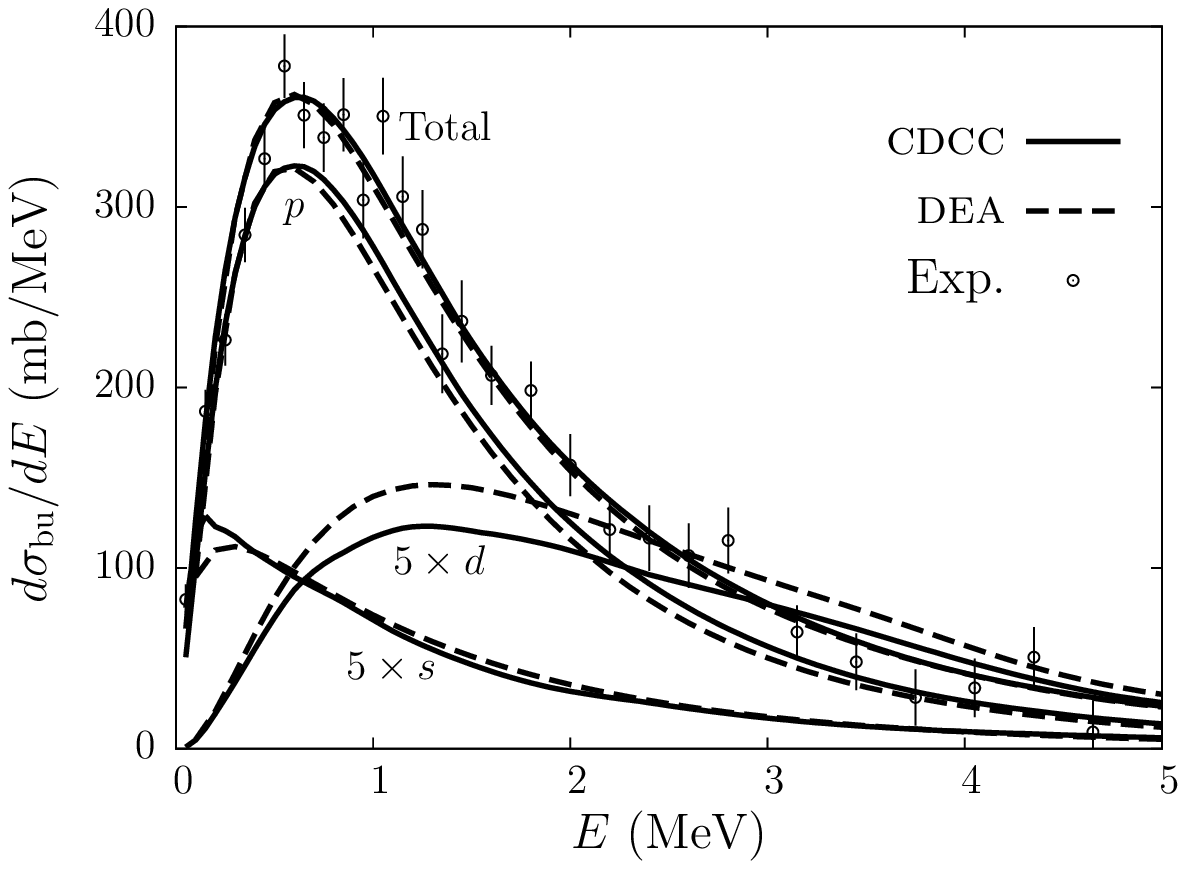,width=4in}
\end{center}
\caption{Relative energy $^{14}$C-n distribution following the breakup of $^{15}$C on Pb at 68 AMeV: comparing the partial wave contributions for CDCC and DEA.}
\label{cdcc-dea68}
\end{figure}

The dynamic eikonal method (DEA) \cite{dea} which is based on the eikonal approximation, has been
developed with the intent of providing a computationally feasible alternative. DEA is non-perturbative, fully quantal
and can equally handle nuclear and Coulomb breakup processes.
Capel {\it et al.} \cite{capel11} compared DEA and CDCC. In Figs. \ref{cdcc-dea20} and \ref{cdcc-dea68}
we show the energy distribution for the breakup of $^{15}$C on Pb at 20 AMeV and 68 A MeV.
While DEA performs well at the higher energy, it produces too large a cross section for the lower energy.
The separate contributions of the various n-$^{14}$C partial waves are included in Figs. \ref{cdcc-dea20}
and \ref{cdcc-dea68}. Focusing on the 20 AMeV case, we see that DEA produces a similar energy distribution for the breakup into p-waves
compared to CDCC. However, it predicts a larger contribution for the even partial waves than CDCC,
which results in a larger total cross section. This discrepancy of DEA and CDCC at lower energy is due to
the fact that DEA does not account for the strong Coulomb deflection properly \cite{capel11}.

The time dependent approach (TD) \cite{td1,td2,td3} is a non-perturbative semiclassical approach, also benchmarked against CDCC \cite{capel11}.
This method is applicable when the breakup process is Coulomb dominated. Contrary to DEA, TD does account for
the Coulomb deflection since it assumes the projectile follows Coulomb trajectories. It is found
that TD provides a reliable prediction of the angle integrated cross sections for both energies 20-70 A MeV.
However, due to its semiclassical nature, TD cannot provide detailed angular distributions.

\subsection{Reducing ambiguities}

Regardless of the important theoretical developments in the description of breakup of loosely bound exotic
system, one large source of uncertainty remains, namely the core-target interaction.
The ratio method \cite{ratio}, inspired by an earlier work on elastic scattering \cite{reb},
was developed with the aim of getting rid of this uncertainty.
Its present form focuses on the breakup of one-neutron halo nuclei: $P+T \rightarrow c+n+T$. 
It was realized that the diffraction pattern in elastic and breakup angular distributions are very similar.
The original idea of Capel {\it et al.} \cite{ratio} was to take the ratio of the breakup cross section and the elastic cross section
to obtain a new observable that depends only on the structure of the loosely bound projectile.
For practical reasons, it was concluded that instead of using only the elastic, it was best to use
a summed cross section:
\beq
\frac{d\sigma_{\rm sum}}{d\Omega}=\frac{d\sigma_{\rm el}}{d\Omega} +\frac{d\sigma_{\rm inel}}{d\Omega}
+\int \frac{d\sigma_{\rm bu}}{dEd\Omega} dE.
\eeqn{e6}
Then the ratio $\frac{d\sigma_{\rm bu}/dEd\Omega}{d\sigma_{\rm sum}/d\Omega}$
provides directly a projectile form factor, independent of the reaction details: 
\beq
|F_{E,0}|^2=\sum_{ljm}\left|\int\phi_{ljm}(E,\ve{r}) \phi_0(\ve{r}) e^{i\ve{Q\cdot r}}d\ve{r}\right|^2.
\eeqn{e4}
Here  $\phi_{ljm}(E)$ is the projectile's scattering wave function  corresponding to $c$-$n$ energy $E$
in partial wave $ljm$ and $\phi_0$ is the projectile's ground state wave function.
The quantum numbers $l$, $j$, and $m$ are, respectively,
the $c$-$n$ orbital angular momentum,
the total angular momentum resulting from the coupling of $l$ with the
neutron spin, and the projection of $j$.

\begin{figure}[t]
\begin{center}
\psfig{file=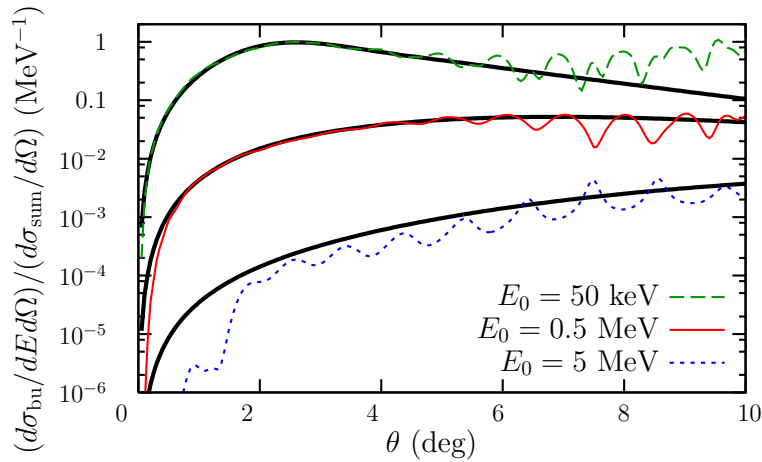,width=4in}
\end{center}
\caption{Sensitivity of the ratio of the breakup and elastic cross section as a function of
the binding energy of the system}
\label{ratio-be}
\end{figure}
In Fig. \ref{ratio-be} we show the strong sensitivity of the form factor to the projectile's neutron
separation energy.
The thick black lines are the results calculated directly from Eq.\ref{e4} while the thin colored lines correspond
to the results obtained from the ratio of the cross sections computed within DEA. The calculated cross sections correspond to the
breakup of $^{11}$Be on $^{12}$C at 68 A MeV, although results on $^{208}$Pb are very similar.
As discussed \cite{ratio}, this new observable can also be used to extract the angular momentum of the ground state
as well as more detailed properties of the projectile's ground state.

\section{Outlook}

In this presentation we cover a number of recent developments pertaining nuclear reactions with rare isotopes.
We focus on transfer reactions and breakup. Present non-perturbative methods for calculating (d,p) cross sections have
been benchmarked and results show that around 10 A MeV, ADWA and CDCC are in agreement with the exact Faddeev calculations. 
Also discussed is the use of DOM as input to transfer calculations, which bring uniformity to the spectroscopic factors
extracted from (d,p) data at different beam energies.

The comparison with exact Faddeev has also shown the limitation of ADWA and CDCC at higher beam energies.
In order to be able to address reactions with heavy nuclei in the future, it is important that new methods be
formulated which can indeed handle the larger Coulomb field and provide the necessary accuracy.

We also include a discussion on the description of breakup reactions of loosely bound nuclei. In this context the
DEA and TD methods have been benchmarked against CDCC and results reveal the level of accuracy to be expected from
these methods in predicting the energy and angular distributions at various beam energies. To reduce the uncertainty
of the calculated breakup cross section from the ambiguities in the optical potential between the core and
the target (by far the largest uncertainty in most cases of interest), the ratio method has been developed.
Here we briefly discuss the main idea behind this method and the type of properties that the new ratio observable can reveal.

In all examples discussed in this presentation, it was assumed that the nucleus under study was pure single particle.
Clearly extensions are needed to be able to handle two-particle states and one-particle mixed configuration. Although there have
been developments in that direction \cite{xcdcc,cdcc4a,cdcc4b}, these are hindered by computational limitations. 
It would be advantageous to the field that simpler reaction models, that have been proven accurate in important
regimes, be extended to include more complex projectile structure.

\medskip

Support from the National Science Foundation grant PHY-0555893 and PHY-096894
and from the U.S. Department of Energy, Division of Nuclear Physics under grants DE-FG52-08NA28552 and FG02-87ER-40316 and the TORUS collaboration DE-SC0004087 are acknowledged.

\bibliographystyle{ws-procs9x6}

\end{document}